\documentclass[aps,prb,amsmath,groupedaddress,letterpaper,showpacs,twocolumn,superscriptaddress]{revtex4}
\usepackage{graphicx,array,tabularx}
\usepackage{bm,amsmath,amssymb,verbatim}

\begin{document}

\title{Sinusoidal electromagnon in $R$MnO$_3$: Indication of anomalous magnetoelectric coupling}
\author{Markku P. V. Stenberg}
\email[]{markku.stenberg@iki.fi}
\affiliation{Department of Physics and Astronomy, University of Victoria,
Victoria, B.C., V8W 3P6, Canada}
\affiliation{Department of Microelectronics and Nanoscience,
Chalmers University of Technology, S-41296 G\"oteborg, Sweden}
\author{Rog\'{e}rio de Sousa}
\email[]{rdesousa@uvic.ca}
\affiliation{Department of Physics and Astronomy, University of Victoria,
Victoria, B.C., V8W 3P6, Canada}

\begin{abstract}
  The optical spectra in the family of multiferroic manganites
  $R$MnO$_3$ is a great puzzle.  Current models can not explain the
  fact that two strong electromagnons are present in the non-collinear
  spin cycloidal phase, with only one electromagnon surviving the
  transition into the collinear spin sinusoidal phase.  We show that
  this is a signature of the presence of anomalous magnetoelectric
  coupling that breaks rotational invariance in spin space and
  generates oscillatory polarization in the ground state.
\end{abstract}
\date{\today}

\maketitle

\section{Introduction}

In multiferroic materials magnetic and electric orders coexist
simultaneously and the coupling between spin and charge degrees of
freedom gives rise to a wide range of magnetoelectric phenomena
\cite{eerenstein06,tokura06,cheong07}. Recent research has centered on
the origin and symmetry of magnetoelectric coupling. The crucial
question is how the coupling between two spins depends on electric
field:
\begin{eqnarray}
H_{\rm{me}} &=& \sum_{nm}\left[J_{nm}({\bf E})\hat{{\bf S}}_{n}\cdot \hat{{\bf S}}_{m} 
+ {\bf D}_{nm}({\bf E})\cdot \hat{{\bf S}}_{n}\times \hat{{\bf S}}_{m}\right.
\nonumber\\ 
& &+ \left.\hat{{\bf S}}_{n}\cdot {\bf A}_{nm}({\bf E})\cdot \hat{{\bf S}}_{m}
\right].
\label{fme}
\end{eqnarray}
Here $\hat{{\bf S}}_n$ and $\hat{{\bf S}}_m$ are spins at lattice
sites $\mathbf{R}_n$ and $\mathbf{R}_m$, and the electric field ${\bf
  E}$ can be either internal, i.e., from the electric polarization in
the material, or external, as is the case of incident light.  The
first two interactions in the right hand side of Eq.~(\ref{fme}) are
well understood.  The first one, exchange interaction $J$, is
electric-field dependent because atomic positions are modulated by
${\bf E}$ (the phenomena of magnetostriction). The second one, the
Dzyaloshinskii-Moriya (DM) interaction, is first order in spin-orbit
coupling and is antisymmetric under spin interchange.
The third and final interaction, the anomalous tensor ${\bf
  A}$, is instead symmetric under spin interchange; it is known to
originate from second order spin-orbit effects \cite{moriya60}, but is
usually believed to be weak or hard to probe. Nevertheless, its
electric field dependence has not been studied.

Most interesting effects take place when one of the coupling
coefficients depends \emph{linearly} on electric field. For instance,
simple models based on electronic \cite{katsura05} or lattice mediated
polarization \cite{sergienko06} predict that the DM vector is
electric-field dependent according to $\bm{D}_{nm}\propto {\bf E}\times
({\bf R}_n - {\bf R}_m)$; this gives rise to the phenomena of
magnetically induced ferroelectricity observed in a large class of
materials, the cycloidal multiferroics \cite{kimura03,katsura05,sergienko06,mostovoy06}. 

In addition, the linear magnetoelectric effect makes magnetic
excitations electrically dipole active. This gives rise to the
electromagnon, the quasiparticle of the multiferroic state
\cite{baryakhtar69,pimenov06,katsura07,takahashi08,aguilar09,kida09,lee09,stenberg09,mochizuki10,rovillain11}.
The observation of electromagnons in optical experiments provide
invaluable clues on the symmetry and magnitude of the magnetoelectric
coupling present in Eq.~(\ref{fme}). Moreover, the ability to launch, detect, and control
magnons electrically also holds promise for novel applications in
information processing \cite{desousaAPL08,rovillain10}.

\begin{figure}
\includegraphics[width=0.5\textwidth]{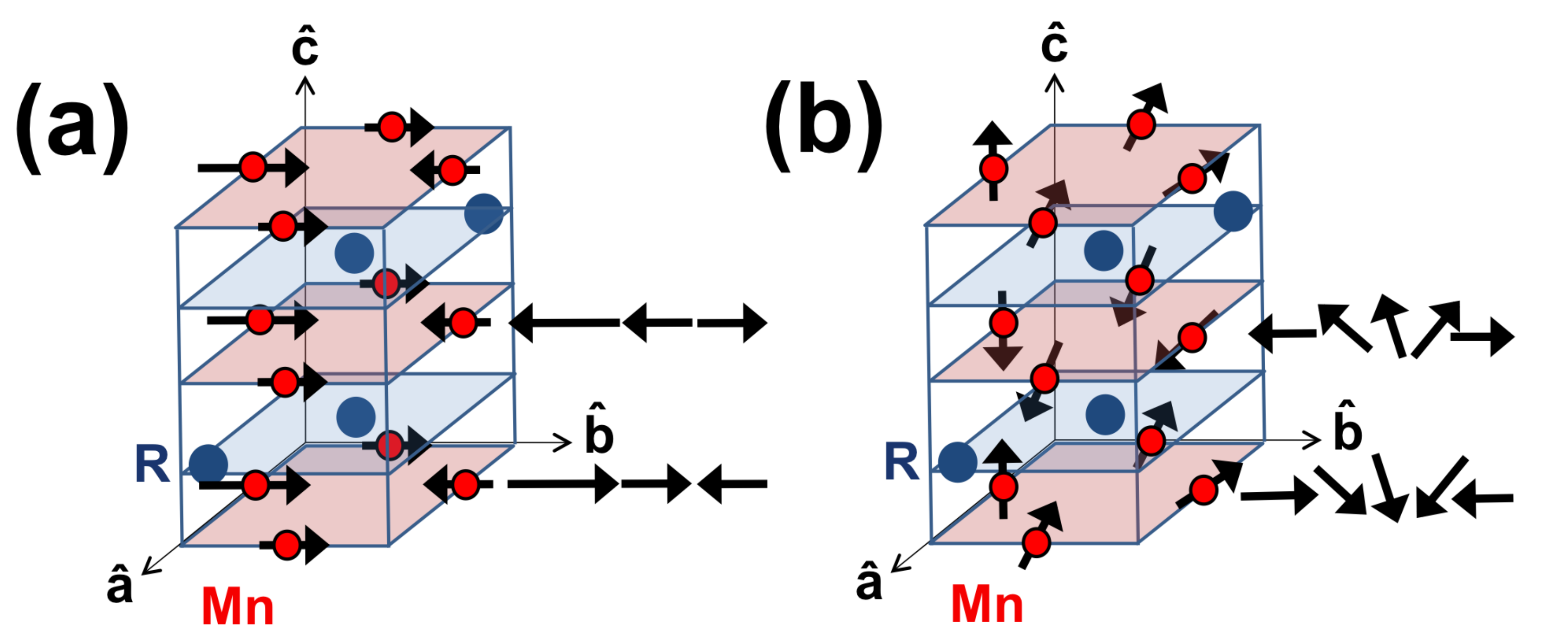}
  \caption{(Color online) Schematics of low temperature phases of $R$MnO$_3$: (a) Collinear sinusoidal phase and (b) Non-collinear cycloidal phase.}
\label{fig1}
\end{figure}

The observation of magnetically-induced ferroelectricity
\cite{kimura03} and electromagnons \cite{pimenov06} in the class of
perovskite manganites $R$MnO$_3$ has made this material the prototype
for studies of strong magnetoelectric effects.  Here $R$ is a
rare-earth ion such as Dy, Tb, Gd or their mixture, {\it e.g.},
Gd$_{x}$Tb$_{1-x}$. In the $R$MnO$_3$ family, spins are typically
ordered with a period incommensurate with the lattice
\cite{kimura03,kida09}. Below the first N\'eel temperature ($T=39$~K in
DyMnO$_3$), the ground state of the Mn spins forms the collinear
sinusoidal density wave depicted in Fig.~\ref{fig1}(a). At even lower
temperature ($19$~K in DyMnO$_3$), another phase transition takes
place where the Mn spins order non-collinearly in the cycloid ground
state shown in Fig.~\ref{fig1}(b).

The detection of electromagnons in the cycloidal phase of $R$MnO$_3$
led to a surprising observation. Optical experiments showed that two
quite strong electromagnons are observed in the cycloidal phase,
provided the electric field of light was directed along the
crystallographic direction $\hat{\mathbf{a}}$
\cite{kida08,aguilar09,lee09}. This remained true even when the cycloid plane 
was flipped, leading to the conclusion that the DM interaction ${\bf D}$ could 
not explain the origin of the observed strong electromagnon resonances (but in 
recent experiments a weak electromagnon resonance consistent with the DM model
was observed \cite{shuvaev10}). This is a surprising conclusion in view of
the fact that the DM interaction is known to be responsible for
ferroelectricity in these materials. 

Optical experiments have also produced a puzzling observation: 
{\it The lower energy electromagnon, unlike the higher energy one, 
survives also in the collinear sinusoidal phase}. This is observed, e.g., 
in DyMnO$_3$ \cite{kida08}, Gd$_{0.7}$Tb$_{0.3}$MnO$_3$ \cite{kida08jpsj},
and Eu$_{1-x}$Y$_x$MnO$_3$ \cite{takahashi09}, but in TbMnO$_3$
no electromagnons are discernible in the sinusoidal phase \cite{takahashi08}.

Currently, there exists a consensus that the high energy electromagnon
originates from magnetostriction, the first term in Eq.~(\ref{fme})
\cite{aguilar09,stenberg09,mochizuki10}.  However, no consensus exists
on the origin of the low energy electromagnon.  Two quite different
models were proposed for its explanation: In [\onlinecite{stenberg09}], we
showed that magnetostriction plus spin-orbit coupling is
able to explain the origin of both electromagnons even when the
cycloid ground state is purely harmonic. In [\onlinecite{mochizuki10}],
Mochizuki {\it et al.}  showed that pure magnetostriction plus cycloid
anharmonicity (without a tensor ${\bf A}$) is able to explain the two
electromagnons of the cycloid phase, suggesting that anharmonicity
plays a vital role (similar results for BiFeO$_3$ were proposed in
[\onlinecite{desousa08b}]). {\it But neither of the two above-mentioned models 
is able to explain the optical activity of the low energy electromagnon in the
sinusoidal phase}.

\section{Model for $R\rm{MnO}_3$}

Here we present a model of electromagnon excitations that can explain
the optical experiments in both the sinusoidal and in the cycloidal
phases. Our model Hamiltonian consists of spin and phonon couplings,
$H=H_{S}+H_{\rm{ph}}+H_{\rm{me}}^{(1)}+H_{\rm{me}}^{(2)}$. Here $H_S$
describes exchange interactions and single-ion anisotropies,
\begin{equation}
H_{S} = \sum_{n,m}J_{n,m}\hat{{\bf S}}_{n}\cdot\hat{{\bf S}}_{m}
+D_a\sum_{n}(\hat{{\bf S}}_{n}\cdot \hat{\bf{a}})^{2}-
D_b\sum_{n}(\hat{{\bf S}}_{n}\cdot \hat{\bf{b}})^{2}.
\label{eq:H_S}
\end{equation}
We assume $D_a>0$ and $D_b>0$, favoring alignment along the
$\hat{\mathbf{b}}$ direction.  The spins are coupled by exchange
interactions $J_{n,m}$, with nearest-neighbor interactions in the ab
plane denoted by $J_0$, next-nearest-neighbor interaction along ${\bf
  \hat{b}}$ denoted by $J_{2b}$, and interaction along the ${\bf
  \hat{c}}$ direction denoted by $J_c$. The interaction $J_0<0$ is
ferromagnetic while $J_{2b}>0$ and $J_c>0$ are both antiferromagnetic. 

At sufficiently low temperatures, provided that the stability
condition $J_{2b} > -J_0/2$ is satisfied, the competition between the
nearest-neighbor ferromagnetic exchange, and the antiferromagnetic
next-nearest-neighbor exchange favors incommensurate spin ordering.
Between the first and second N\'eel temperatures 
the spins order in a sinusoidal density wave,
\begin{equation}
\bm{S}_{0}({\bf R},T) =
\pm S(T)\cos({\bf Q}\cdot {\bf R}+\phi)\hat{\bf{b}},
\label{groundstate}
\end{equation}
with $S(T)$ a temperature dependent amplitude [see Eq. (3.14) in 
Ref.~\onlinecite{nagamiya67} for its dependence on model parameters].
The magnitude of the sinusoidal wave vector $Q$ is given by $\cos(Qb/2) =
-J_0/(2 J_{2b})$. The upper sign in Eq.~(\ref{groundstate})
corresponds to ab layer spins with the integer c-coordinate, while the
lower sign applies to spins in the neighboring ab layers a distance
$c/2$ above and below them.

Our phonon Hamiltonian is
\begin{equation}
H_{\rm{ph}}=\frac{1}{2}m^{*}\sum_n \left(\dot{\bm{x}}_{n}^{2}
+\omega_{0}^{2} \bm{x}_{n}^{2}\right)
-e^{*}\sum_n \bm{x}_n \cdot \bm{E},
\label{hph}
\end{equation}
where $\omega_0$ is the (bare) phonon frequency, $m^{*}$ is the
effective mass, $\bm{x}_{n}$ is the relative displacement between
anions and cations in the $n$th unit cell, $e^{*}$ is the Born charge,
and $\bm{E}$ is the electric field of light.

We divide the linear magnetoelectric couplings in our model into two
separate terms, $H_{me}^{(1)}$ and $H_{me}^{(2)}$.
The first interaction,
\begin{eqnarray}
H_{\rm{me}}^{(1)} &=&
e^{*}\sum_{n}x_{n}^{a}[
g_{c}(\hat{S}_{1,n}^{c}-\hat{S}_{1,n+b}^{c})(\hat{S}_{2,n}^{c}
+\hat{S}_{2,n+a}^{c})\nonumber\\& &+g_{b}(\hat{S}_{1,n}^{b}
-\hat{S}_{1,n+b}^{b})(\hat{S}_{2,n}^{b}
+\hat{S}_{2,n+a}^{b})\nonumber\\& &+(1\rightarrow 3, 2\rightarrow 4)],
\label{eq:linear_pbnm1}
\end{eqnarray}
does not give rise to electromagnons in a collinear ground state, but
is necessary to explain the origin of the low frequency electromagnon
in the cycloidal phase \cite{stenberg09}. 
The second interaction is instead given by
\begin{eqnarray}
H_{\rm{me}}^{(2)} &=&
e^{*}\sum_{n}x_{n}^{a}\{[
g_{bc}(\hat{S}_{1,n}^{b}-\hat{S}_{1,n+b}^{b})(\hat{S}_{2,n}^{c}
+\hat{S}_{2,n+a}^{c})
\nonumber\\& &+g_{ab}(\hat{S}_{1,n}^{a}
-\hat{S}_{1,n+b}^{a})(\hat{S}_{2,n}^{b}
+\hat{S}_{2,n+a}^{b})
+(1\leftrightarrow 2)]\nonumber\\& &+(1\rightarrow 3, 2\rightarrow 4)\},
\label{eq:linear_pbnm2}
\end{eqnarray}
where $g_{ab}$ and $g_{bc}$ are coupling constants that can be obtained 
by microscopic calculation (e.g., using density functional theory). 
Like Eq.~(\ref{eq:linear_pbnm1}), this 
spin-symmetric interaction is also invariant under the $Pbnm$ space-group 
operations of $R$MnO$_3$, and is therefore consistent with lattice symmetry. 
Both interactions represent anomalous magnetoelectric coupling, with
particular forms of the anomalous tensor $\bm{A}$ [Eq.~(\ref{fme})].
A generalization of Moriya's theory \cite{moriya60} to allow for
magnetostriction effects shows that such interactions can originate
from cross-coupling between spin-orbit and magnetostriction effects.
However, a full microscopic theory is still needed to confirm this
expectation.

\section{Electromagnon spectra}

We adopt the molecular field approximation and expand the Hamiltonian
$H$ by keeping only terms quadratic in the fluctuation operators,
e.g., $\delta\hat{S}^2_c$, $\delta\hat{S}_a\delta P_a$, $\delta P_a^{2}$, etc.  
We parametrize the spin excitations $\delta
\hat{\bm{S}}=\hat{\bm{S}}-\bm{S}_{0}$ by $\delta \hat{\bm{S}}_{i,n}= \hat{s}^{a}_{i,n} \hat{\mathbf{a}}
\pm  \hat{s}^{c}_{i,n}\hat{\mathbf{c}}$, and compute the equations of motion 
using the canonical commutation relations,
$[\hat{s}^c_{j,n},\hat{s}^a_{k,m}]=i\delta_{jk}\delta_{nm}\hat{S}^{b}_{k}$.
In addition, we also adopted the random phase approximation (RPA),
i.e., we made the substitution $\hat{S}^{b}_{k}\rightarrow
\langle\hat{S}^{b}_{k}\rangle=\bm{S}_{0}(\bm{R}_k,T)\cdot
\hat{b}$ in the commutator above. Such an approximation is expected to
hold when the fluctuation effects are not too large (i.e., we are
sufficiently far from the N\'{e}el temperature).

\begin{figure}
\includegraphics[width=0.5\textwidth]{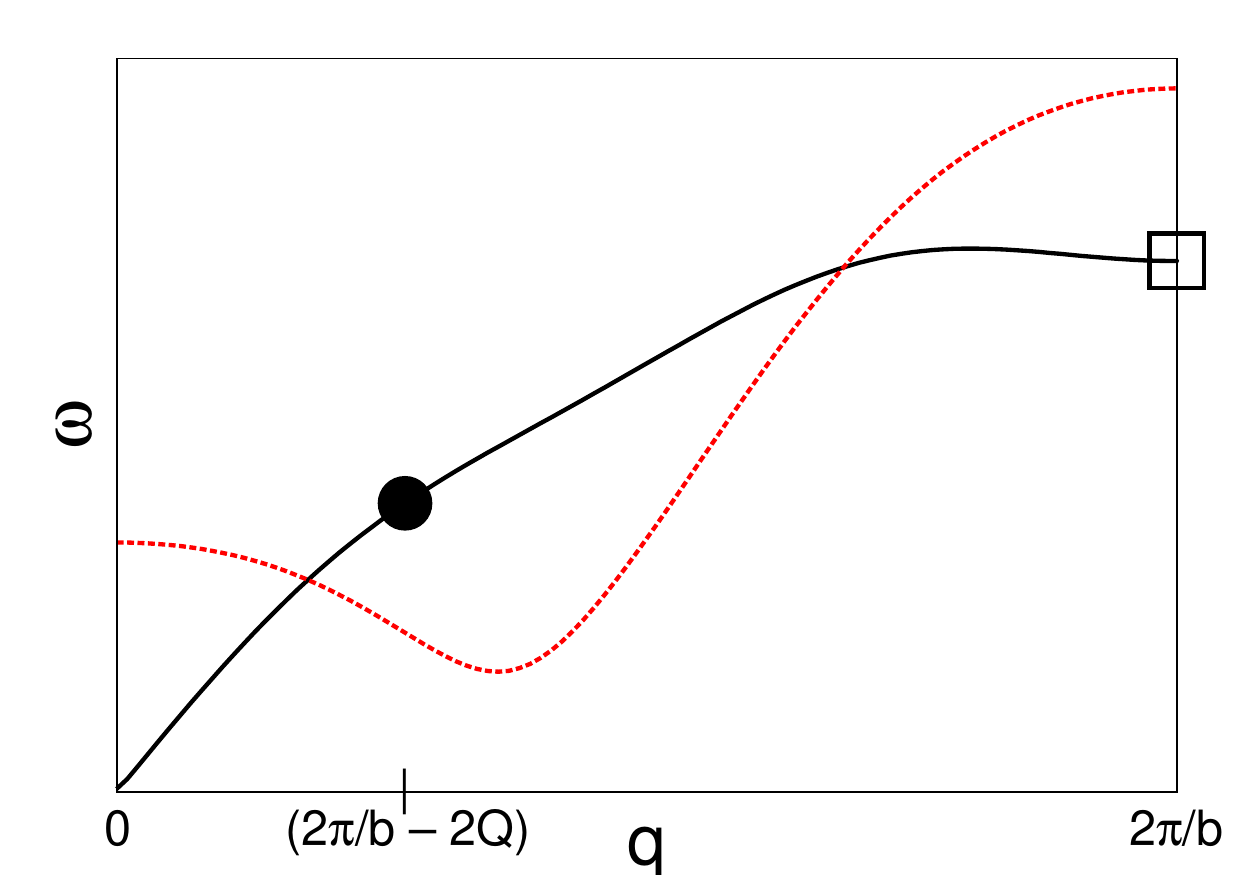}
  \caption{(Color online) Typical dispersion curves for magnon wavevector ${\bf q}$ along $\hat{\bf{b}}$
in the sinusoidal phase: cyclon (black solid) and extra-cyclon (red dashed).
In the sinusoidal state, only the low-energy electromagnon (filled circle) 
is activated, through $H_{me}^{(2)}$ [Eq.~(\ref{eq:linear_pbnm2})]. 
In the cycloidal state, $H_{me}^{(1)}$ [Eq.~(\ref{eq:linear_pbnm1})]
activates also the zone-edge electromagnon (hollow square).}
\label{fig2}
\end{figure}

After some manipulation the coupled equations of motion for spins and polarization is given by,
\begin{subequations}
\begin{eqnarray}
\left(\omega^{2}-\Omega_{\rm{C},q}^{2}\right)
\left(s^{\alpha}_{1q}+s^{\alpha}_{2q}+s^{\alpha}_{3q}+s^{\alpha}_{4q}\right)&=&\Omega_{\rm{C},q}\Gamma_q,
\label{cyclon}
\\
\left(\omega^{2}-\Omega_{\rm{C},q+k_0}^{2}\right)
\left(s^{\alpha}_{1q}-s^{\alpha}_{2q}+s^{\alpha}_{3q}-s^{\alpha}_{4q}\right)&=&\Omega_{\rm{C},q+k_0}\Gamma_{q+k_0},
\label{zecyclon}
\nonumber\\
\\
\left(\omega^{2}-\Omega_{\rm{EC},q}^{2}\right)
\left(s^{\alpha}_{1q}+s^{\alpha}_{2q}-s^{\alpha}_{3q}-s^{\alpha}_{4q}\right)&=&0,
\label{extracyclon}
\\
\left(\omega^{2}-\Omega_{\rm{EC},q+k_0}^{2}\right)
\left(s^{\alpha}_{1q}-s^{\alpha}_{2q}-s^{\alpha}_{3q}+s^{\alpha}_{4q}\right)&=&0,
\label{zeextracyclon}
\end{eqnarray}
\end{subequations}
where $\alpha = a,c$.  Here $s^{\alpha}_{iq}$ is the momentum
representation of the spin fluctuation $s^{\alpha}_{in}$. Equations
(\ref{cyclon}) and (\ref{zecyclon}) are related by a shift in momentum
space, ${\bm q} \leftrightarrow \bm{q}+\bm{k}_0$ where $k_0=2\pi/b$ is
the Brillouin zone-edge for magnons.  Such a relationship corresponds
to the fact that ``anti-phase'' fluctuations of neighboring spins with
wave vector $q$ are equivalent to ``in-phase'' fluctuations at $q +
2\pi/b$. They describe a mode here referred to as a cyclon, with
dispersion $\Omega_{C,q}$. Similarly, Eqs.~{(\ref{extracyclon}) and
  (\ref{zeextracyclon})} share the same momentum shift relationship,
but describe a different mode referred to as an extra-cyclon. The
cyclon and extra-cyclon dispersions are shown in Fig.~\ref{fig2}.
We note that the cyclon has a gap proportional to $D_b$, while
the extra-cyclon has a gap proportional to $(2J_c+D_b)$.

Equations (\ref{cyclon}) and (\ref{zecyclon}) show that $H_{me}^{(2)}$
couples only a single electromagnon, the cyclon at $q=k_0-2Q$, to the
polar phonon. This takes place through dynamic magnetoelectric
coupling $\Omega_{\rm{C},q}\Gamma_q$ with
\begin{eqnarray}
\Gamma_q&=&\Gamma_q^{ab}-\Gamma_q^{bc},\quad \Gamma_q^{bc}=\Gamma_q^{ab}(g_{ab}\rightarrow g_{bc}),\\
\Gamma_q^{ab}&=&\frac{g_{ab}S(T)^{2}v_0\sin{\left(\frac{Qb}{2}\right)}\delta
P_0^a}{\hbar}\left[e^{-2i\phi}\delta_{q-k_0+2Q}\right.\nonumber\\
&&-e^{2i\phi}\delta_{q-k_0-2Q}+\left.e^{-2i\phi}\delta_{q+k_0+2Q}-e^{2i\phi}\delta_{q+k_0-2Q}\right].\nonumber\\
\end{eqnarray}

Optical experiments such as transmissivity or reflectivity probe the frequency dependence of the
dielectric function $\epsilon(\omega)$. After a linear response calculation we obtain
\begin{equation}
\varepsilon(\omega)
=\frac{\mathcal{S}_{em}}{\Omega_{C,k_0-2Q}^2-\Delta^2-\omega^2}+\frac{\mathcal{S}_{ph}}{\omega_{0}^2
+\Delta^2-\omega^2}+\varepsilon_\infty.
\end{equation}
Hence $\epsilon(\omega)$ can be written as two Lorentzians, with poles
at downshifted magnon and upshifted phonon frequencies. The pole at the magnon
frequency shows that the cyclon at $q=k_0-2Q$ is actually an
electromagnon, with spectral weight given by
\begin{equation}
\mathcal{S}_{em}=\frac{4\pi\chi_0 \omega_0^2 \Delta^2}{\omega^2_0-\Omega_{C,k_0-2Q}^2}.
\end{equation}
Here $\chi_0=e^{*2}/(m^{*}v_0\omega_{0}^{2})$ is the zero-frequency
susceptibility, with $v_0$ the unit cell volume. The frequency shift 
$\Delta$ is calculated to be
\begin{equation}
\Delta^2\approx \frac{
S(T)^2e^{*2}(g_{bc}-g_{ab})^2\tan^2\left(\frac{Qb}{2}\right)\Omega_{C,k_0-2Q}^2}{2m^{*}(\omega_0^2-\Omega_{C,k_0-2Q}^2)\left[\sin^4
\left(\frac{Qb}{2}\right)+\cos^4\left(\frac{Qb}{2}\right)\right]J_{2b}}, 
\end{equation}
apart from smaller terms of order $(g_{ab}-g_{bc})g_{ab}$.
Since the magnitude of the frequency shift $\Delta$ is the same for the
magnon and the phonon, we confirm the oscillator strength sum rule
$\mathcal{S}_{em}+\mathcal{S}_{ph}=4\pi\chi_0\omega_0^2$. 

\section{Additional consequence of anomalous magnetoelectric interaction: Incommensurate oscillatory polarization}

In addition to the sinusoidal electromagnon, the couplings described
by Eqs.~(\ref{eq:linear_pbnm1})~and~(\ref{eq:linear_pbnm2}) have an
important observational consequence: They lead to an incommensurate oscillatory
polarization (IOP) with wavevector $2Q$ \cite{stenberg09}. Minimizing
Eqs.~(\ref{hph})--(\ref{eq:linear_pbnm2}) with respect to the
polar phonon displacement ${\bf x}_n$ and plugging in the cycloidal
spin order we get
\begin{eqnarray}
\frac{e^*{\bf x}_{n}}{v_0}&=&4\chi_0 S^2 \sin{\left(\frac{Qb}{2}\right)}\left\{(g_b-g_c)\sin{[Qb(2n+1)]}\right.\nonumber\\
&&\left. -g_{bc}\cos{[Qb(2n+1)]}+g_{bc}\right\}\hat{\bf{a}}.
\label{piopcycloidal}
\end{eqnarray}
Note how $g_{bc}$ generates a combination of static and oscillatory polarization along the $\hat{\bf{a}}$ direction. 

When the system goes into the sinusoidal phase, this polarization changes discontinously to
\begin{equation}
\frac{e^*{\bf x}_{n}}{v_0}=4 \chi_0 S(T)^2
\sin{\left(\frac{Qb}{2}\right)} g_b\sin{[(2n+1)Qb]}\hat{\bf{a}}.
\label{PIOP}
\end{equation}
Such an oscillatory polarization can be detected by X-ray scattering.
Indeed, Kimura {\it et al.} detected an oxygen oscillation with
wavevector $2Q$ in both the cycloidal and sinusoidal phases [see blue
dots in Fig.~1(c) of [\onlinecite{kimura03}]]. 
Just like our prediction, the X-ray intensity in [\onlinecite{kimura03}] showed an apparent discontinuity 
in oxygen displacements in the transition from cycloidal to sinusoidal phase.

Table I shows how a combination of optical and X-ray scattering
experiments are capable of measuring the magnetoelectric coupling
constants individually, and even cross-check some of them. For
instance, for DyMnO$_3$, we obtain from the measured electromagnon
spectral weights \cite{kida08} and X-ray diffraction intensities
\cite{kimura03prb} in the sinusoidal and cycloidal phases the values
of $g_b\sim 170\ {\rm erg/(cm\ esu)}$, $g_c\sim -40\ {\rm erg/(cm\
  esu)}$, $g_{ab}\sim 100\ {\rm erg/(cm\ esu)}$ $\gg$ $g_{bc}$. 
In TbMnO$_3$, the sinusoidal electromagnon could not be observed
experimentally \cite{takahashi08}. This indicates that $g_{ab},g_{bc}$
$\ll$ $g_{b}$, $g_{c}$, i.e., $H^{(2)}_{\rm{me}}$ is much weaker than
$H^{(1)}_{\rm{me}}$ in TbMnO$_3$.

\begin{table} 
\begin{center} 
\begin{tabular}{c ||c | c } 
 & Electromagnons (Far-IR) &  Atomic disp. (X-ray) \\ 
\hline\hline
Cycloidal &  
\begin{tabular}{c}
$(g_b+g_c)$\\
$(g_b-g_c)$\\
\end{tabular}
& 
\begin{tabular}{c}
$(g_b-g_c)$\\
$g_{bc}$\\
\end{tabular}
\\
\hline
Sinusoidal & $(g_{bc}-g_{ab})g_{ab}$ & $g_b$\\
\end{tabular} 
\caption{This table relates our predictions for optical and X-ray
  experiments to the anomalous magnetoelectric coupling parameters
  $g_\alpha$ and $g_{\alpha \beta}$ introduced in this paper
  ($\alpha,\beta=a,b,c$ are crystallographic directions). The first
  column refers to measurements of electromagnon spectral weight using
  far-IR optical experiments, and the second column refers to the
  measurement of magnetically induced lattice distortions using X-ray
  spectroscopy.  Each experiment has different magnetoelectric
  signatures, depending on whether the ground state is cycloidal or
  sinusoidal.  All parameters can be measured individually, and in
  addition, parameters $g_b$ and $g_c$ can be cross-checked.}
\label{table1} 
\end{center} 
\end{table} 

\section{Discussion and conclusion}

We now consider the justification of our model and other possibilities
for the activation of the electromagnon in the collinear sinusoidal
phase. First, we note that the presence of DM interaction in principle
also predicts an electromagnon in the collinear sinusoidal state
\cite{chupis07,senff08}. However, this scenario is ruled out
by the experiments where the electromagnon is activated with ${\bm E}$
along $\hat{{\bm a}}$ only. In addition to the DM interaction, there
are no other linear magnetoelectric couplings anti-symmetric under the
exchange of spins that would be allowed by lattice symmetry. 

The anharmonic cycloid model of Mochizuki {\it et al.}
\cite{mochizuki10} would give rise to no electromagnon activity in the
sinusoidal phase. One possibility for the activation of sinusoidal
electromagnons in this scenario would be to include additional
single-ion anisotropy so that the spins in the ground state are tilted
off the $\hat{b}$ axis \cite{mochizuki09}. However, in this case {\it
  both high and low energy electromagnons get activated}. We found no
scenario where deformation of the sinusoidal ground state activates
the low-energy electromagnon without activating the high-energy one.

Concerning other possible symmetry-allowed magnetoelectric
interactions, we note that the other couplings quadratic in spin do
not couple electric field linearly to magnons. More specifically,
terms of the form $x^a_n S^a_i S^a_j$, $x^a_n S^c_i S^c_j$, and $x^a_n
S^a_i S^c_j$ lead to contributions that are third order in fluctuation
operators, and the term $x^a_n S^b_i S^b_j$ does not couple
polarization to magnons. Hence in the collinear sinusoidal state only
the couplings linear in $S^b$ considered in the present work
[Eq.~(\ref{eq:linear_pbnm2})] can be responsible for the
electromagnons in collinear sinusoidal state.

In conclusion, we showed that anomalous magnetoelectric coupling gives
a natural explanation for the origin of electromagnons in both the
cycloidal and sinusoidal phases of $R$MnO$_3$. It remains an open
question to study, e.g., through ab-initio methods, the microscopic
mechanism of anomalous magnetoelectric coupling.

We acknowledge support from NSERC discovery.

\end{document}